\documentclass[a4paper,english,10pt,twoside]{article}
\usepackage[english]{babel}
\usepackage[latin1]{inputenc}
\usepackage{amsmath}
\usepackage{graphicx}

\usepackage{fancyhdr}
\fancypagestyle{plain}{%
  \fancyhf{}%
  \fancyhead[C]{\small\it{Twenty years of giant exoplanets - Proceedings of the Haute Provence Observatory Colloquium, 5-9 October 2015\\ Edited by I. Boisse, O. Demangeon, F. Bouchy \& L. Arnold} }
  \fancyfoot[C]{\normalsize \thepage}%
}

\newcommand{\nc}   {\newcommand}
\nc{\dsfrac}[2]{{\displaystyle\frac{#1}{#2}}}

\nc{\ergs}  {\mathrm{erg}/\mathrm{gm\;s}}
\nc{\ergc}  {{\rm erg}~{\rm cm}^{-3}}
\nc{\ergcs} {{\rm erg}~{\rm cm}^{-2}~{\rm s}^{-1}}
\nc{\ms}   {\ensuremath{{\rm m}~{\rm s}^{-1}}}

\nc{\aap}   {A\&A}
\nc{\aaps}   {A\&AS}
\nc{\apj}   {ApJ}
\nc{\apjl}   {ApJL}
\nc{\apjs}   {ApJS}
\nc{\nat}   {Nature}
\nc{\mnras}   {MNRAS}
\nc{\aapr} {A\&ARv}
\nc{\pasp} {PASP}
\usepackage{txfonts}
\usepackage[squaren,cdot]{SIunits}
\usepackage{tabularx}
\usepackage[breaklinks=true]{hyperref} 
\usepackage{natbib,twoopt}
\bibpunct{(}{)}{;}{a}{}{,} 
\makeatletter
\newcommandtwoopt{\citeads}[3][][]{\href{http://adsabs.harvard.edu/abs/#3}%
{\def\hyper@linkstart##1##2{}%
\let\hyper@linkend\@empty\citealp[#1][#2]{#3}}}
\newcommandtwoopt{\citepads}[3][][]{\href{http://adsabs.harvard.edu/abs/#3}%
{\def\hyper@linkstart##1##2{}%
\let\hyper@linkend\@empty\citep[#1][#2]{#3}}}
\newcommandtwoopt{\citetads}[3][][]{\href{http://adsabs.harvard.edu/abs/#3}%
{\def\hyper@linkstart##1##2{}%
\let\hyper@linkend\@empty\citet[#1][#2]{#3}}}
\newcommandtwoopt{\citeyearads}[3][][]%
{\href{http://adsabs.harvard.edu/abs/#3}
{\def\hyper@linkstart##1##2{}%
\let\hyper@linkend\@empty\citeyear[#1][#2]{#3}}}
\makeatother

\makeatletter 

\def\@maketitle{%
  \vskip 2em%
  \begin{center}%
  \let \footnote \thanks
    {\LARGE\textbf \@title \par}%
    \vskip 1.5em%
    {\normalsize
      \lineskip .5em%
      \begin{tabular}[t]{c}%
        \@author
      \end{tabular}\par}%
    \vskip 1em%
    {\normalsize \@date}%
  \end{center}%
  \par
  \vskip 1.5em}
\makeatother

\newcommand{\affil}[1]{\small{\hskip-0.55cm #1}}
\newcommand{\acknowledgments}[1]{\small{ \vskip3mm \hskip-0.55cm {Acknowledgments: #1}}}

\begin{document}

\setcounter{page}{1}  

\title{Twenty Years of Precise Radial Velocities at Keck and Lick
  Observatories} 
\author{ Jason T.\ Wright$^1$}
\date{Talk delivered 4 October 2015} 
\maketitle
\affil{$^1$Center for Exoplanets and Habitable Worlds and Department
  of Astronomy and Astrophysics \\ 525 Davey Laboratory, The
Pennsylvania State University, University Park, PA 16802 \\ email:
{\texttt astrowright@gmail.com}
}

\vskip1cm

\begin{abstract}
The precise radial velocity survey at Keck Observatory began over 20 years ago. Its survey
of thousands of stars now has the time baseline to be sensitive to planets with decade-long
orbits, including Jupiter analogs. I present several newly-finished orbital solutions for
long-period giant planets. Although hot Jupiters are generally ``lonely'' (i.e. they are not
part of multiplanet systems), those that are not appear to often have giant companions at
5 AU or beyond.  I present two of the highest period-ratios among planets in a two-planet
system, and some of the longest orbital periods ever measured for
exoplanets.  In many cases, combining Keck radial velocities from those from other long-term surveys at
Lick Observatory, McDonald Observatory, HARPS, and, of course,
OHP spectrographs, produces superior orbital fits, constraining both period and eccentricity better than
could be possible wiith any single set alone.  Stellar magnetic
activity cycles can masquerade as long-period planets.  In most
cases this effect is very small, but a loud minority of stars,
including, apparently, HD 154345, show very strong RV-activity correlations.
\end{abstract}

\section{The Lick Observatory Planet Search}

\begin{figure}[hp]  
  \centering
  \includegraphics[width=5in]{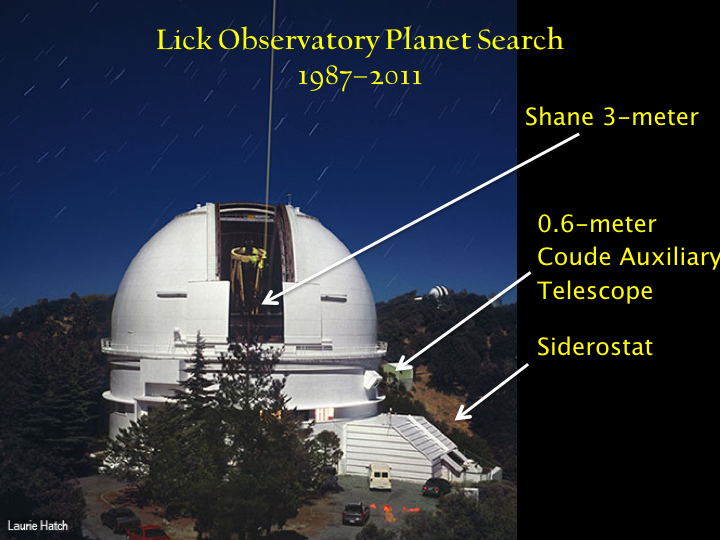}
  \includegraphics[width=3in]{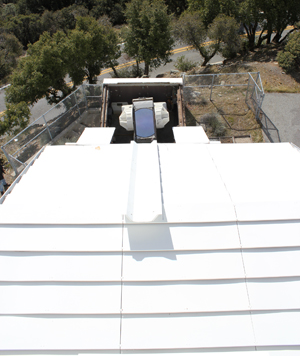}
  \caption{{\it Top:} The Shane 3-m building. Starlight striking the siderostat traveled up to a port in
    the side of the building, where it struck a flat mirror before
    heading to the CAT.  The Hamilton Spectrograph is below ground
    level, beneath the siderostat shed. Photograph by Laurie
    Hatch. {\it Bottom:} View of the siderostat from near the port in
    3-m building. }
  \label{building}
\end{figure}

The precise Doppler planet search at Lick Observatory near San Jose,
California, USA ran from 1987--2011.  It made use of the Hamilton
optical echelle spectrograph in the coud\'e 
room of the Shane 120-inch telescope building, built by Steve Vogt
\citep{Vogt87}. On nights that the Shane 
120-inch telescope was used with other instruments, the Hamilton
spectrograph was often fed via the Coud\'e Auxiliary Telescope (the ``CAT''), a
0.6-meter telescope within the dome that received starlight via a
siderostat outside the building, above the coud\'e room (Figure~\ref{building}).   Together,
these two telescopes allowed bright stars to be monitored on virtually
any clear night.

\begin{figure}[hp]
\centering
\includegraphics[width=4.5in]{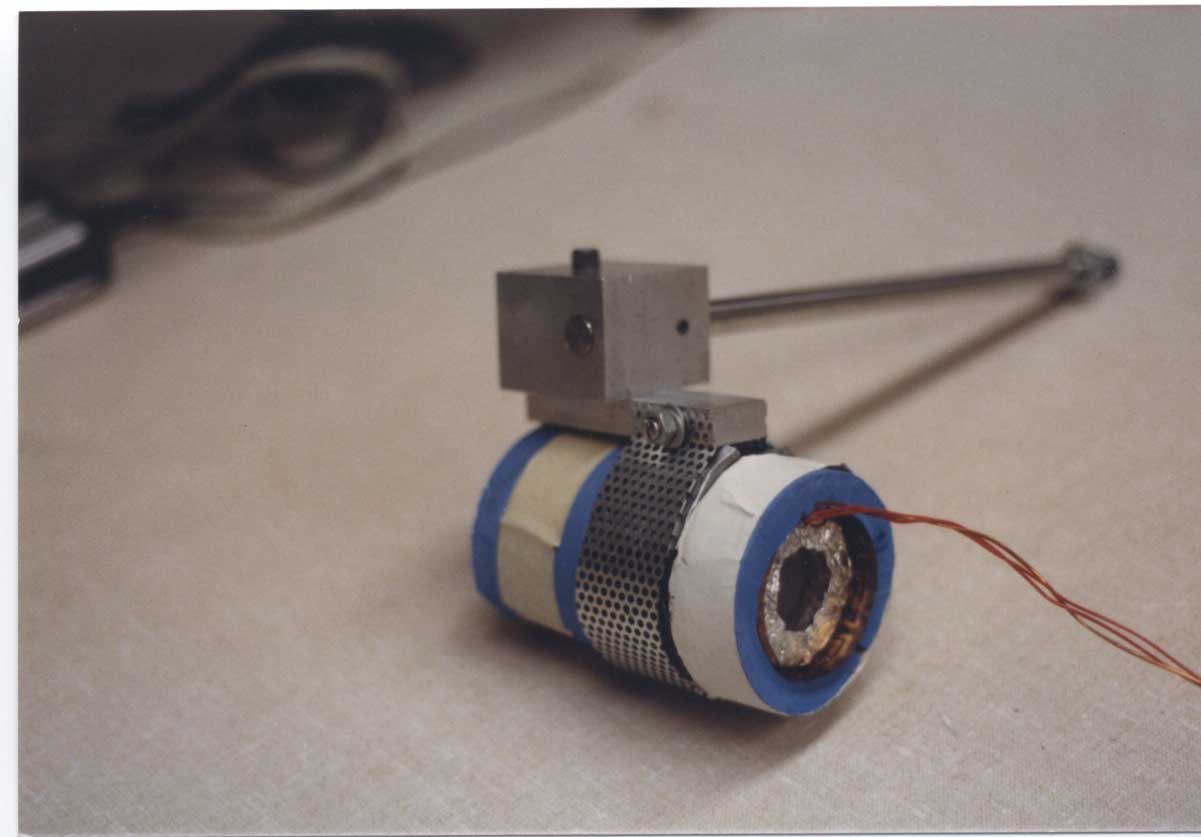}
\includegraphics[width=3in]{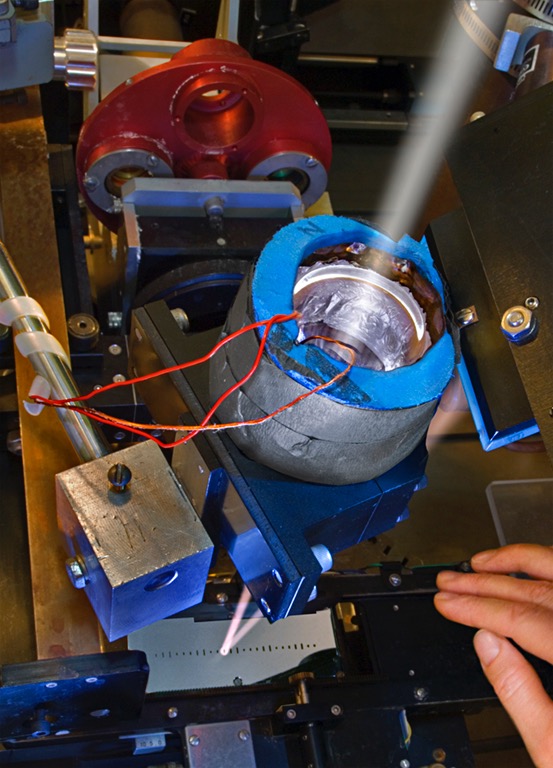}
\caption{{\it Top:} The original Lick iodine cell for Lick, designed by Paul Butler and Geoff
Marcy. {\it Bottom:} The Lick cell in
position before the slit plate in the slit room of the Hamilton
Spectrograph.  Photograph by Laurie Hatch. }
\label{cell}
\end{figure}

The Hamilton Spectrograph was slit-fed, subject to large temperature
variations, and not stable at a level that would allow for
long-term precise Doppler work without special efforts.  Such efforts
by \citet{Marcy92} came in the form of an iodine absorption cell (Figure~\ref{cell}),
following pioneering work by
\citet{Campbell79} using an HF absorption cell, itself inspired by Roger Griffin's
suggestion that clever exploitation of telluric lines would enable 10
\ms\ Doppler precision \citep{Griffin73}.  The use of iodine as an
ideal absorption medium was the suggestion of Robert
Howard, of the Carnegie Institute of Washington, inspired by
\citet{Beckers77} (also note the contemporaneous efforts of
\citet{Libbrecht88} and \citet{Cochran90}). 

\begin{figure}[hp]
\centering
\includegraphics[width=4.5in]{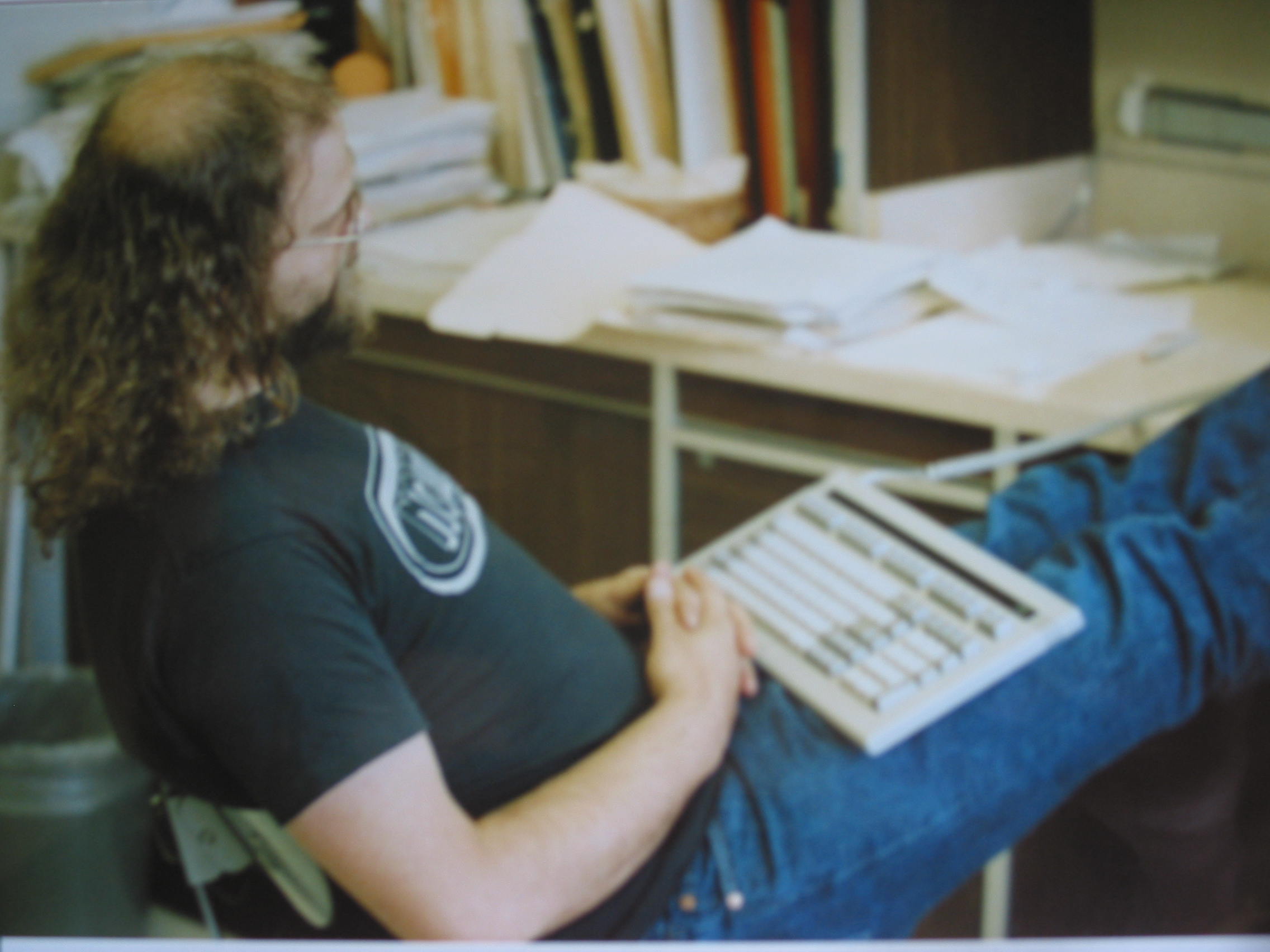}
\includegraphics[width=4.5in]{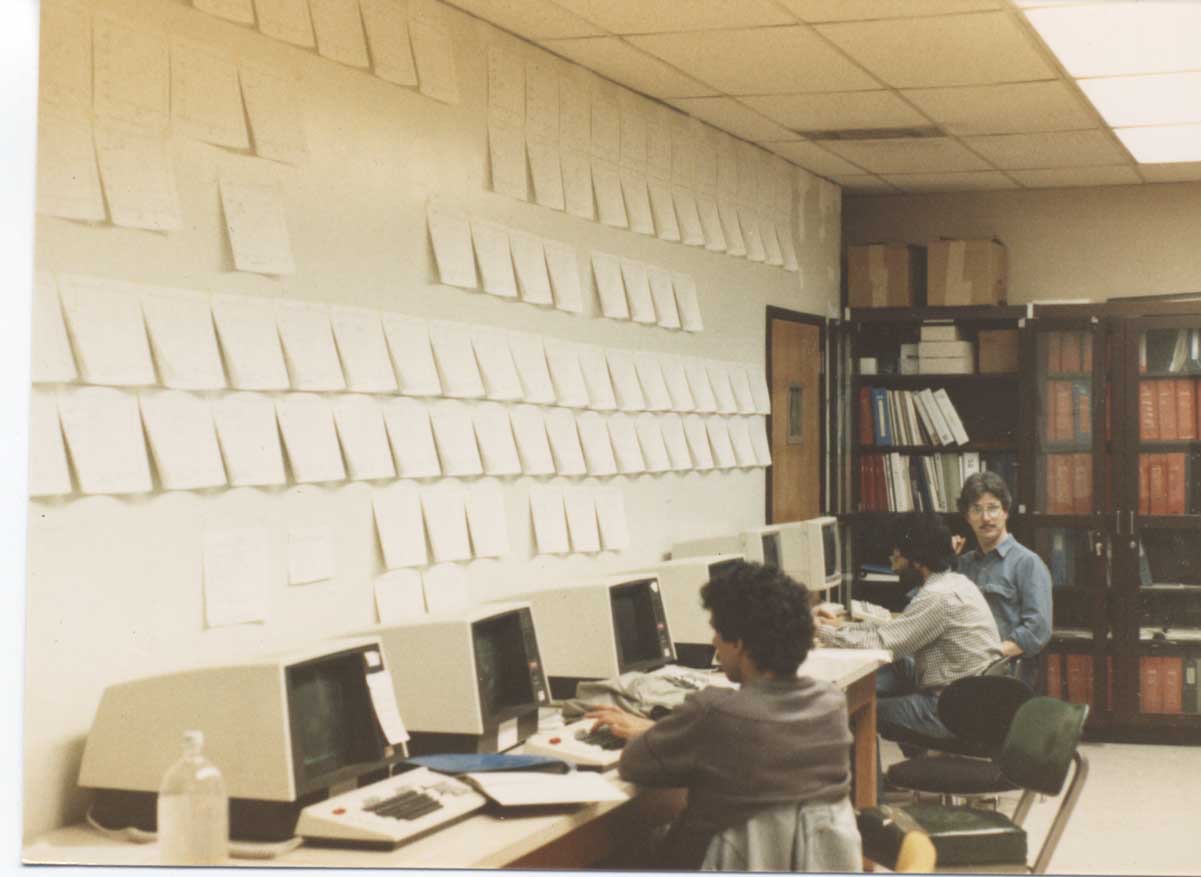}
\caption{Early Doppler work at San Francisco State University.  {\it
    Top:} Paul Butler circa 1988.  Butler's work with Geoff Marcy constructing the
  iodine cell and using its complex absorption features to model the
  instrumental profile allowed the Lick Planet Search to achieve 3
  \ms\ r.m.s. variations on some stars \citep{Butler96b}.  {\it
    Bottom:} The Doppler Lab, with diagnostic plots of the Doppler code
  lining the wall.
}
\label{SFSU}
\end{figure}

The Hamilton echelle used a prism cross-disperser, providing broad
wavelength coverage, and originally used an 800x800 CCD (later upgraded to 2048x2048).  This
combination allowed the extremely complex (and unresolved) iodine
absorption features to be modeled numerically by computer on a
pixel-by-pixel basis, and used to solve for the instrumental profile
of the spectrograph \citep{Valenti95}. Much of this work was done at San
Francisco State University by Geoff Marcy and Paul Butler, and their
collaborators, including Jeff Valenti at CU Boulder (see Figure~\ref{SFSU}).

The work at San Francisco State University by Marcy, Butler, and
others eventually led to Doppler precisions below 10 \ms, including
r.m.s.\ velocity variations for some stars as low as 3 \ms
\citep{Butler96b}.  Shortly after the revolutionary announcement of
the discovery of 51 Peg {\it b} by \citet{Mayor_queloz},
\citet{Marcy_51peg} confirmed the discovery, and they parleyed the ensuing attention to
their program into access to the computing resources they needed to
perform the modeling calculations necessary to thoroughly analyze
their data.  The Lick Planet Search would go on to announce the next nine exoplanet
discoveries, including the first multiplanet system
\citep{Butler_upsand}.

\begin{figure}[hp]
\centering
\includegraphics[width=4in]{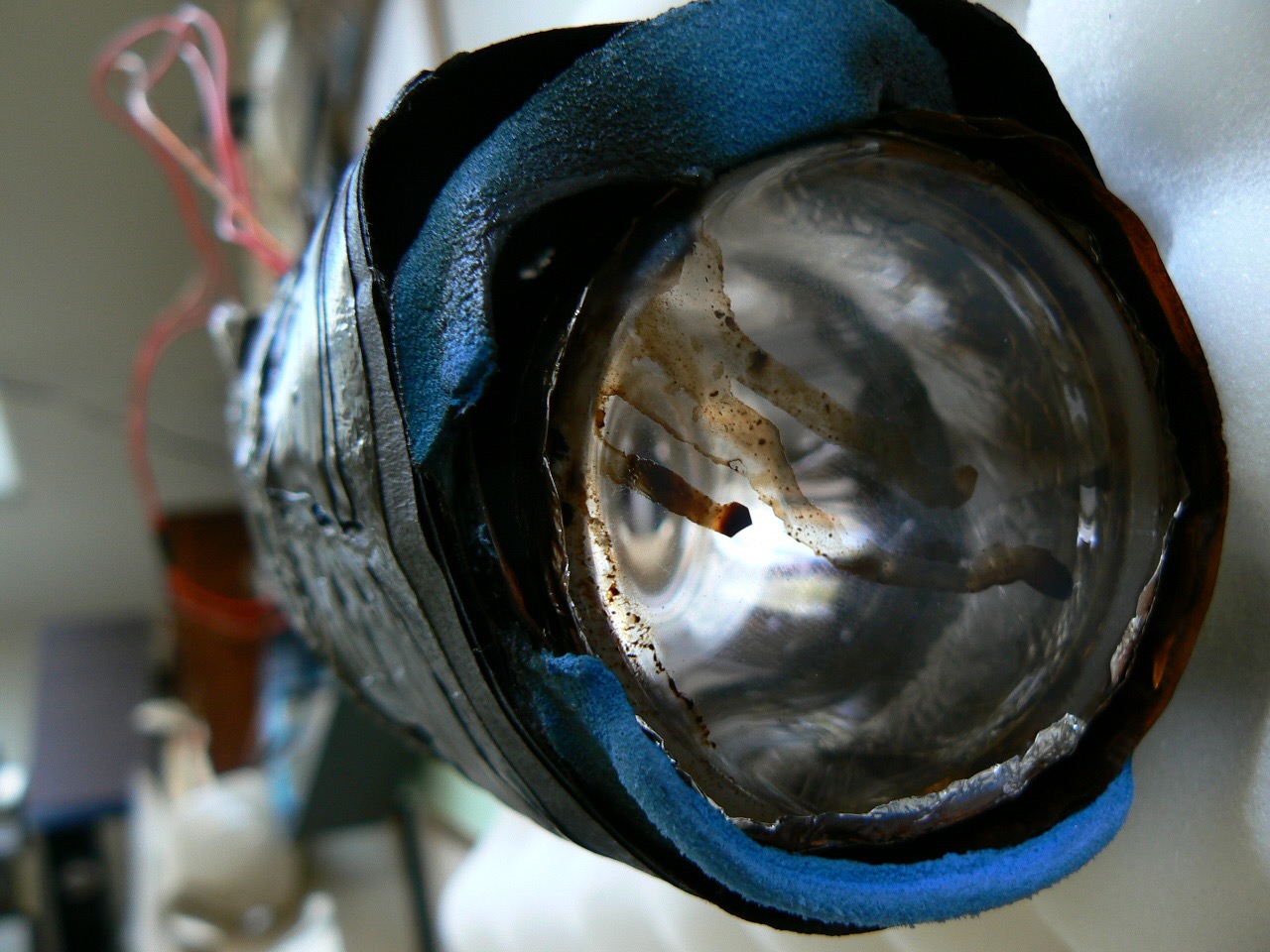}
\caption{The Lick iodine cell after the temperature controller
  malfunction that caused it to badly overheat, effectively ending the
  Lick Planet Search in 2011.}
\label{badcell}
\end{figure}

The Lick Planet Search spanned nearly 25 years, and included several
upgrades to the spectrograph and detector.  In 2011, an error in the temperature controller for the iodine
cell caused the insulation to overheat, badly damaging the cell and
altering its transmission properties (see Figure~\ref{badcell}).  This, combined with the
superior Doppler precision of other facilities, small aperture of the
CAT, and dwindling support for Lick Observatory operations led to the
decision to end the Lick Planet Search.

In total, the Lick Planet Search produced precise radial velocity time series
from over 14,000 observations of 386 stars.  The entire radial
velocity archive, and a more detailed description of the project was published by \citet{Fischer14}.

\section{The Keck Observatory Planet Search}

\begin{figure}[hp]
 \centering
 \includegraphics[width=2.5in]{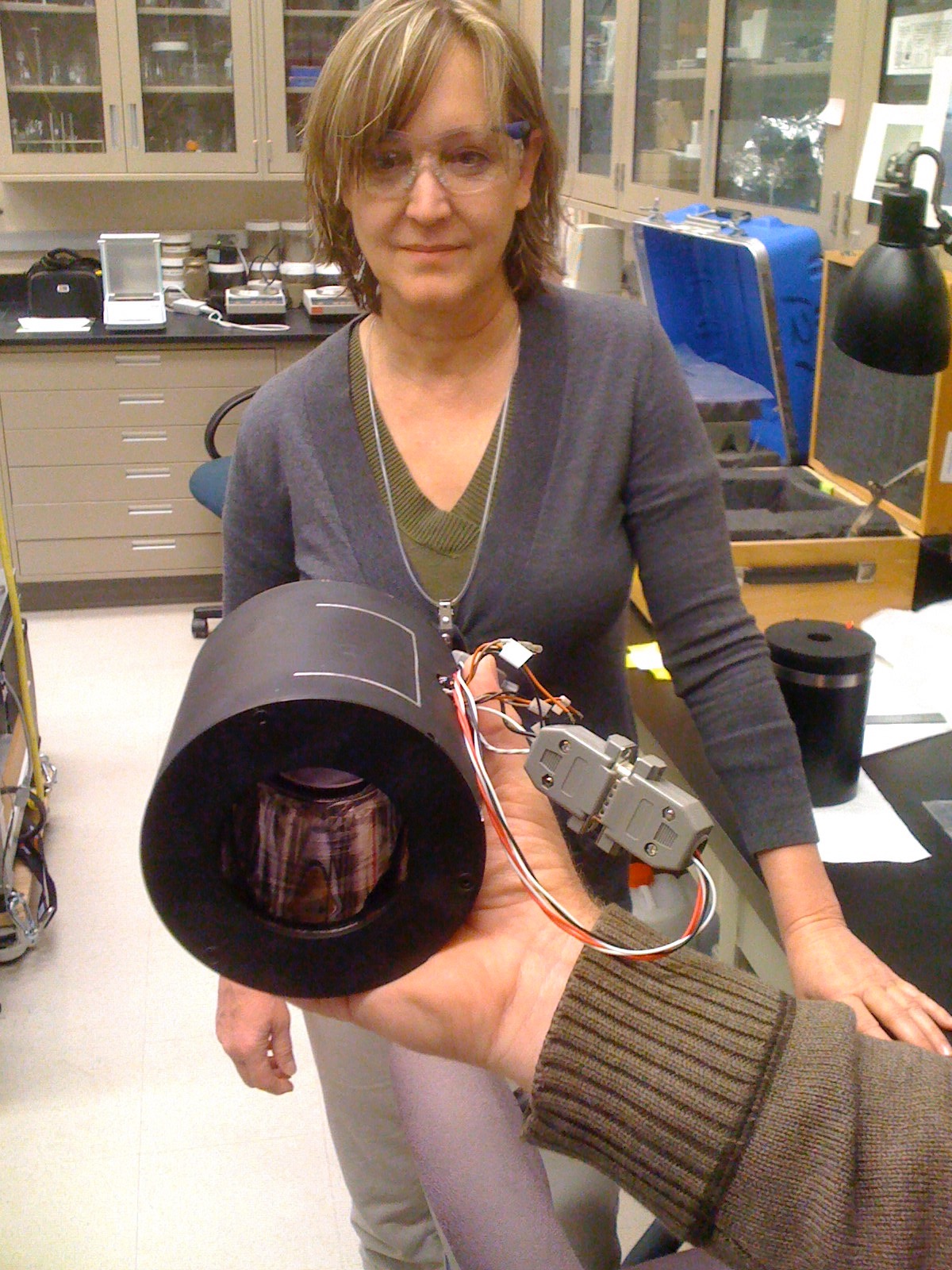}
 \includegraphics[width=2.5in]{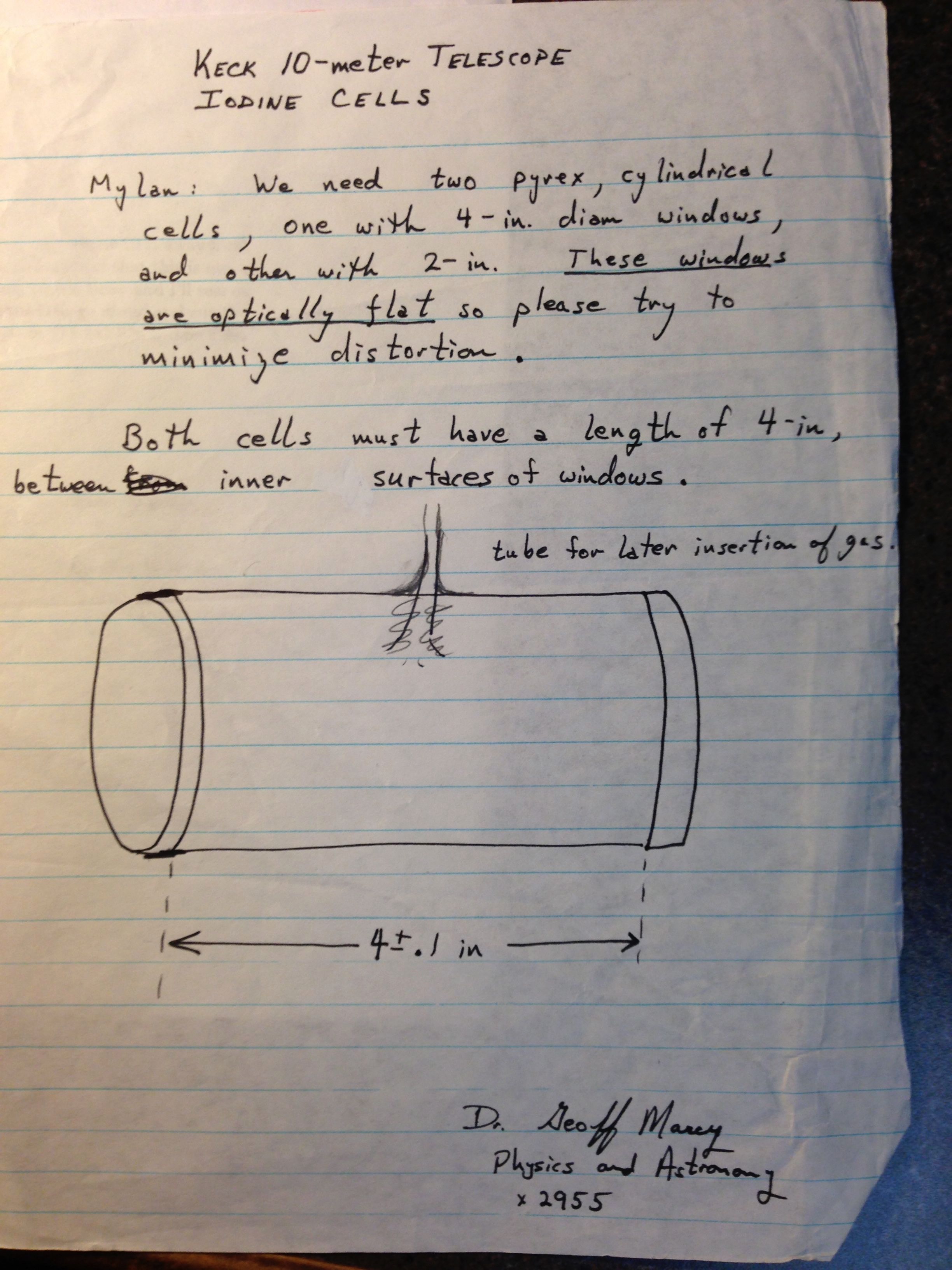}
 \caption{{\it Left}: Debra Fischer and the HIRES iodine cell during
   calibration with a Fourier Transform Spectrograph. {\it Right:}
   Notes by Geoff Marcy for Mylan Healy, a glassblower at San Francisco State
   University, for construction of the Keck iodine cell.}
 \label{KeckCell}
 \end{figure}

\begin{figure}[hp]
 \centering
 \includegraphics[width=4in]{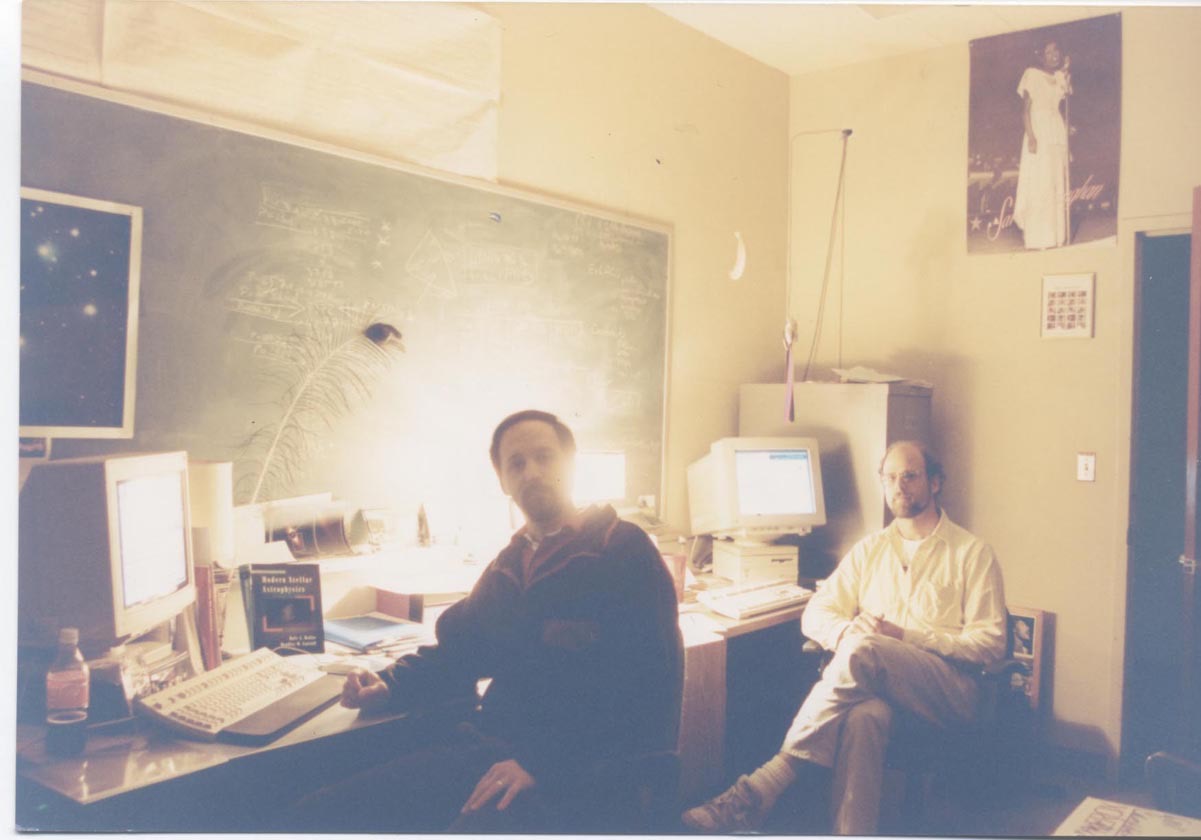}
 \caption{Paul Butler and Geoff Marcy at UC Berkeley c. 1994}
 \label{berkeley}
 \end{figure}

The construction of HIRES at Keck Observatory by \citet{Vogt94}
allowed the Lick Planet Search to move to Keck Observatory.  A new
iodine cell was constructed for HIRES (Figure~\ref{KeckCell}), and the Doppler pipeline was
applied to this new telescope.  Around this time, Geoff Marcy and Paul
Butler began to strengthen their ties with collaborators at UC
Berkeley, especially Gibor Basri, that had been helping with access to
Lick and Keck Observatories and computational facilities.  Butler served as a visiting research
fellow at Berkeley until 1997 and Marcy as adjunct faculty
(Figure~\ref{berkeley}). In 1999, Marcy moved to Berkeley, joined by
postdoctoral fellow Debra Fischer, and in the same year 
Butler joined the Department of Terrestrial Magnetism at the Carnegie
Institute of Washington as a staff scientist.

The iodine planet search at Keck Observatory has been ongoing since 1994. One benefit of the 21+ years of observation at a
single site with better than 3 \ms\ precision, is the ability to
monitor for long-period planets.  Except for a detector upgrade in 2004
that improved precision (and created a break in RV continuity, necessitating that a small RV offset be
calculated for each star at that date) the data set is uniform and
continuous.  As such, many long-period planets, including Jupiter
analogs, have emerged.  

The first true Jupiter analog announced was HD 154345 {\it b}
\citet{Wright08}, an apparent 1 Jupiter-mass planet in a circular,
10-year orbit around a G star.  \citet{Wright08} noted that both
photometry and the activity levels of HD 154345 were consistent with a
magnetic activity cycle, similar to the 11-year solar cycle, and were
well correlated with the precise radial velocities.  Those authors
ruled out spurious RV {\it induced} by this activity cycle by noting
that activity cycles are common among the G and K stars in the
California Planet Survey sample, and that very few of them show
correlated RV variations at the 10 \ms\ level, like HD 154345.
Indeed, the closest spectral match to HD 154345 in the CPS sample,
$\sigma$ Draconis is the ``quietest'' star in the CPS sample and the
quintessential RV stable star, and it has an even higher activity level and
more vigorous magnetic cycle that HD 154345.

\begin{figure}[hp]
\includegraphics[width=5.5in]{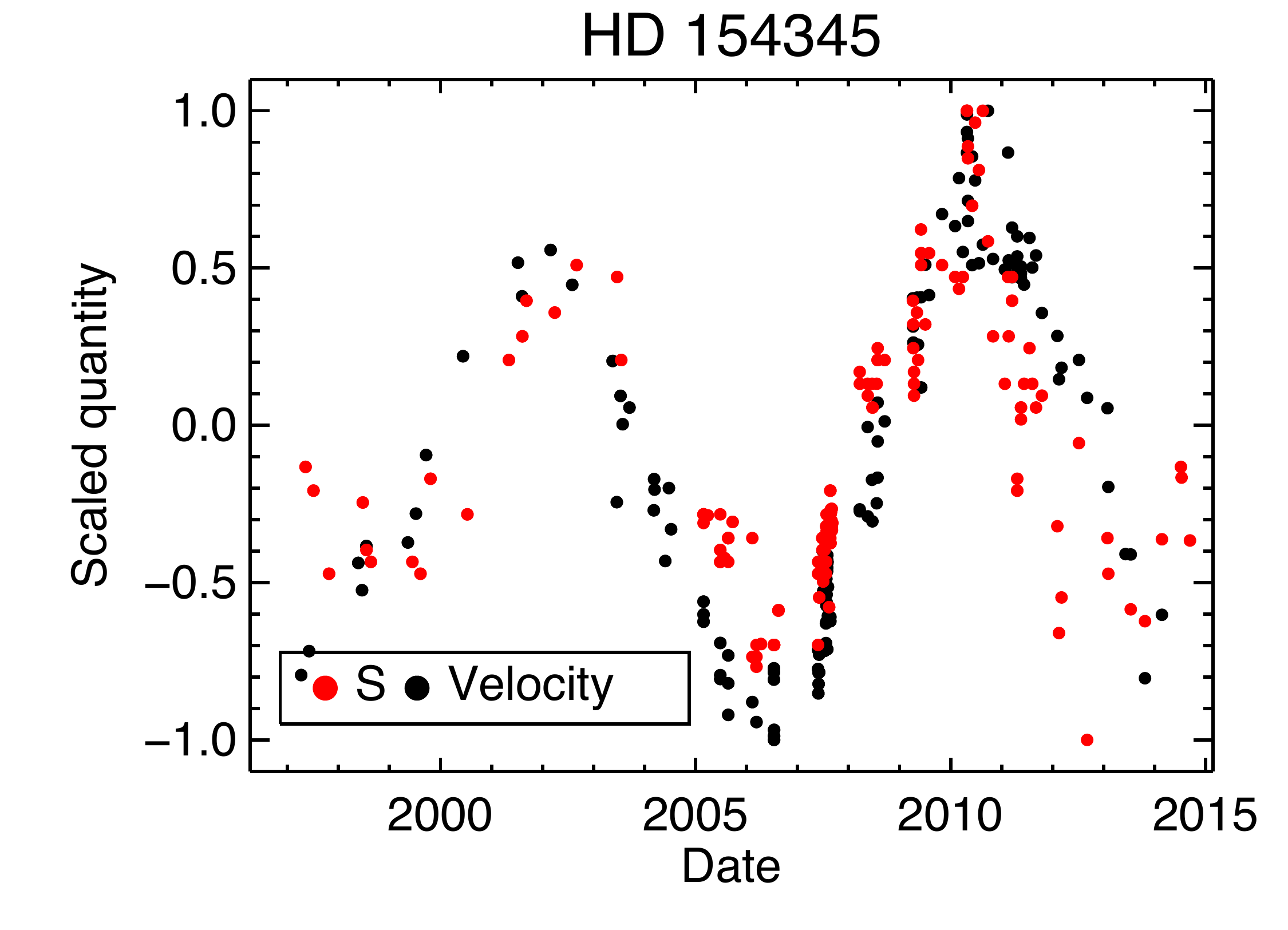}
\caption{Activity levels measured by the Ca {\sc ii} H \& K emission
  lines and radial velocities, shifted and scaled to emphasize their
  correlation.  There is a discontinuity in both data sets in 2004
  because of the detector change.  The sinusoidal signal announced by
  \citet{Wright08} may not be due to a real planet, but rather a rare, but apparently not
  unique, activity-induced, spurious radial velocity shift.}
\label{154345}
\end{figure}

However, continued monitoring of HD 15345 since 2008 has shown that
the activity cycle continues to be well correlated with the RVs,
amplifying the coincidence (see Figure~\ref{154345}).  Similar correlations among a small number
of other stars of similar spectral type have begun to cast doubt on
the planetary hypothesis for HD 154345, and raised a new question
about finding Jupiter analogs: why do a small number of stars appear
to show strong RV-activity correlations, at the 10 \ms\ level and
above, while the vast majority show little, if any such correlation?

\begin{figure}[hp]
\centering
\includegraphics[width=4.5in]{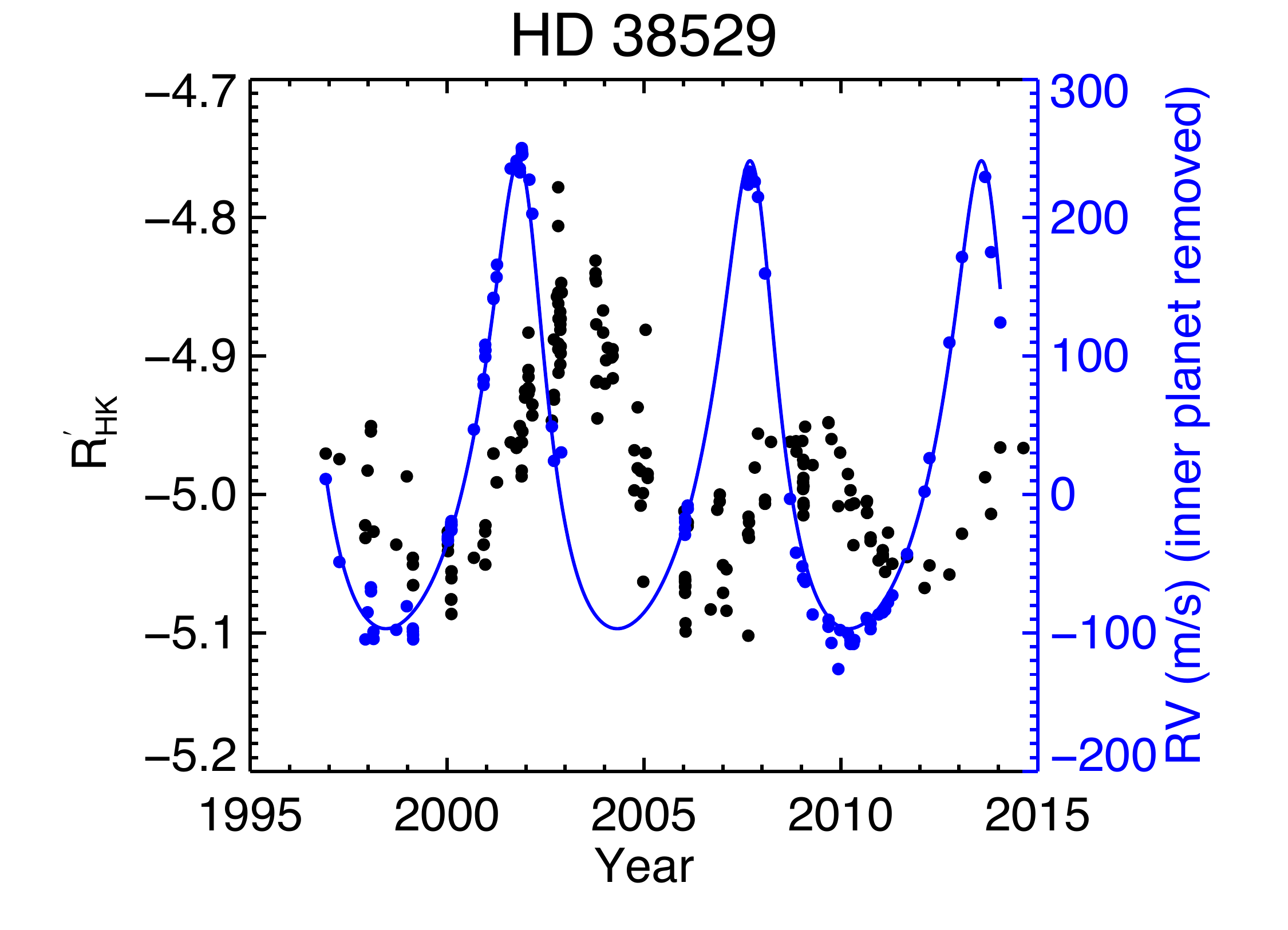}
\includegraphics[width=4.5in]{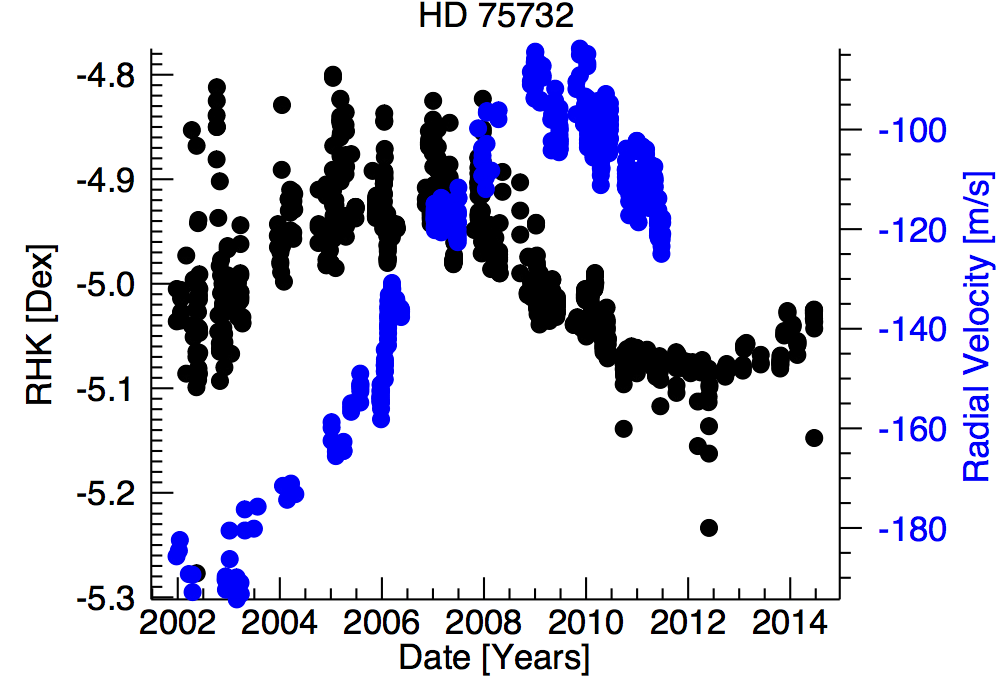}
\caption{Two examples of activity cycles in stars hosting long-period
  planets, illustrating how the effects are easily distinguished in
  most stars. {\it Top:} The long-period planet HD 38529 {\it c} (the
  signal of the short period planet {\it b} has been removed) shows a
  similar period to the activity cycle of the star, but shifted in
  phase, and with a consistent shape, unlike the cycle which has a
  varying strength.  {\it Bottom:} The outer planet of the 55 Cnc
  system (inner components removed) similarly induces RV variations
  with a similar period but different shape and phase from the
  activity variations. Figures by Jacob Brown.}
\label{cycles}
\end{figure}

Long-period giant planets with higher masses are more unambiguous,
because their RV larger amplitudes make them easier to distinguish
from the generally noisier activity-induced RV variations, especially
when they have Keplerian profiles characteristic of eccentricity.
Indeed, despite concerns that activity cycles would make long-period
planets hard to detect, contemporaneous observations of activity
levels make distinguishing them rather easy (when the periods and
phases do not coincide, at least).  Figure~\ref{cycles} shows two
cases when similar-period activity and RV variations have arisen, but
there is little reason to suspect a physical connection.

The contemporaneous planet search efforts with Lick, HET/HRS, ELODIE, SOPHIE, and HIRES
provide an opportunity to merge data sets, providing independent
measures of the behavior of stars between instrument changes.  This is
especially important for long-period planets, whose measured eccentricities,
amplitudes, and orbital periods may be covariant with an unknown
offset introduced by changes to the instruments.

\begin{figure}[hp]
\centering
\includegraphics[width=4.5in]{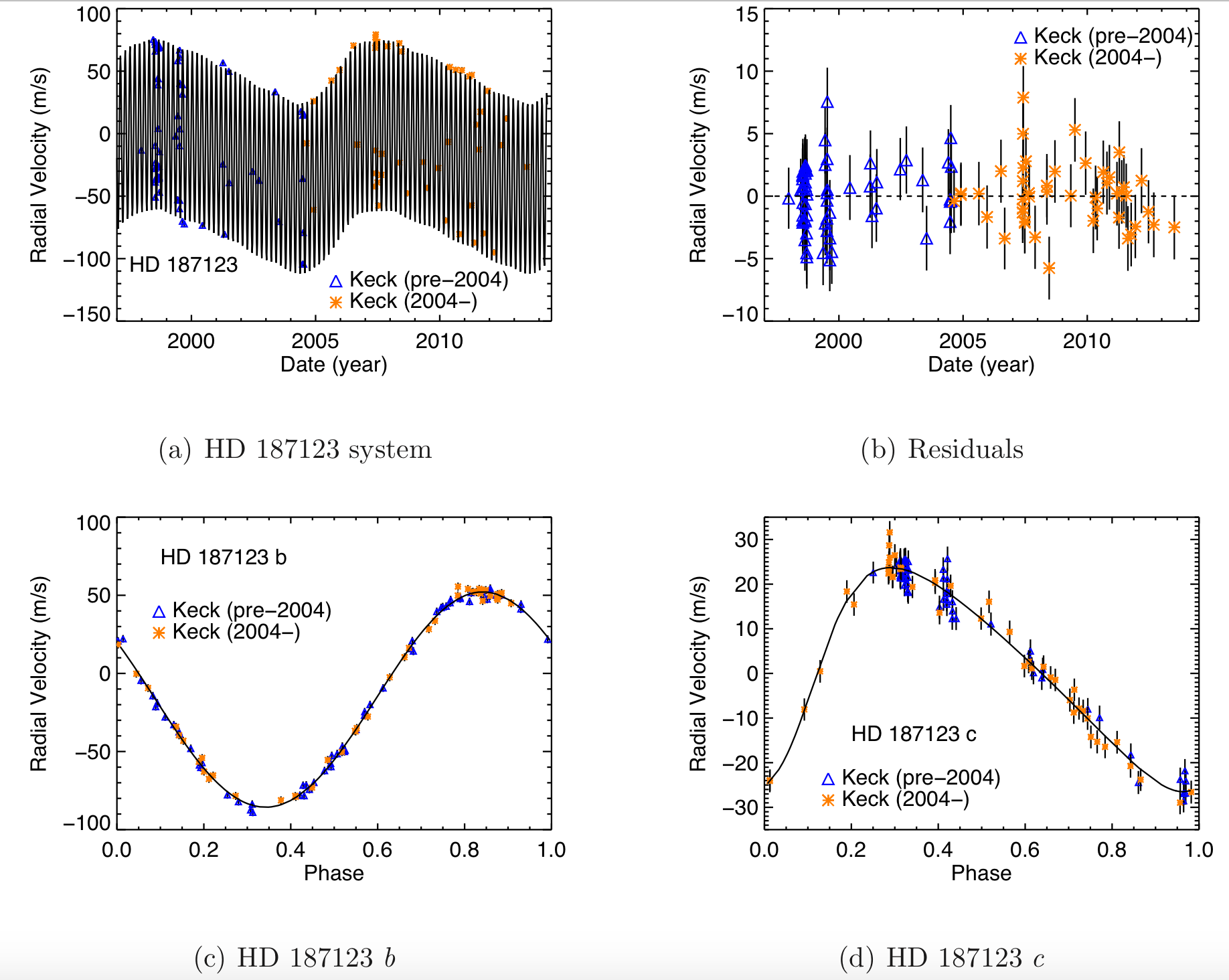}
\includegraphics[width=4.5in]{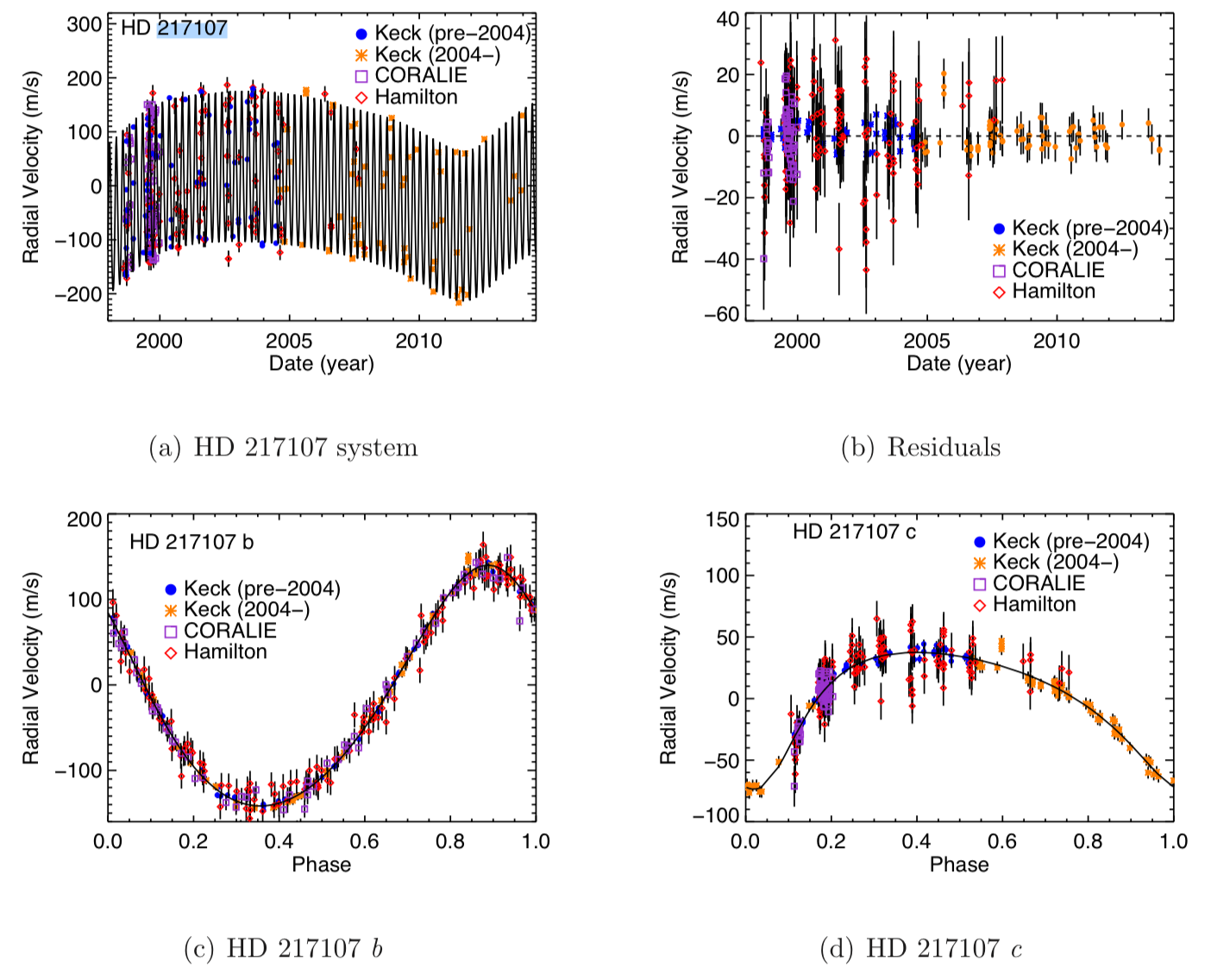}
\caption{Two examples of ``cold friends'' of hot
  Jupiters. \citet{Feng15} used a combination of published CORALIE and
  Lick data with new Keck data to calculate good orbits for the outer
  components of the HD 187123 and HD 217107 systems.  The period
  ratios in these systems are the largest known ($> 1000$), and the period
  ratio for HD 187123 is roughly a factor of 2 greater than the largest ratio for
  planets in the Solar System.  Figures from \citet{Feng15}.}
\label{coldfriends}
\end{figure}

\citet{Wang12} made good use of both HRS/HET and HIRES observations to find HD 37605
{\it c}, an outer super-Jupiter with an 8-year, circular orbit.
\citet{Feng15} used published data from Lick, HET, ELODIE, CORALIE,
and HARPS, combined with new HIRES data, to calculated good orbital
periods for  new and updated orbits for several long-period giant planets, including HD 187123 {\it c} and HD 217107
{\it c}.  These two planets orbit stars already known to host hot
Jupiters, and have the largest orbital period ratios
with respect to their inner planets (over 1000) of any known system,
including the Solar System (see Figure~\ref{coldfriends}).

The emergence of these long-period
companions to hot-Jupiter hosting systems reveals that, while hot
Jupiters may not be ``lonely'' in general, when they do have
companions, these ``friends'' are often
``cold'', and keep their distance.

\acknowledgments{I thank Geoff Marcy for many of the photographs used
  in my talk, and Jacob Brown for preparing Figure~\ref{cycles}. I thank the conference organizers for
inviting me and producing such a successful event. I thank Laurie
Hatch for her work photographing Lick Observatory, and for allowing us
to use her photography in this article.}



\bibliographystyle{aa} 


\end{document}